# Multimodal Autonomous Last-Mile Delivery System Design and Application

***Farah Samouh***
*Laboratory of Innovations in Transportation (LiTrans)*
*Ryerson University*
*Toronto, Canada*
fsamouh@ryerson.ca

***Veronica Gluza***
Laboratory of *Innovations in Transportation (LiTrans)*
*Ryerson University*
*Toronto, Canada*
vgluza@ryerson.ca

***Shadi Djavadian***
Ford Motor Company, Greenfield Labs,
Palo Alto, California
sdjavadi@ford.com

***Seyed Mehdi Meshkani***
Laboratory of *Innovations in Transportation (LiTrans)*
*Ryerson University*
*Toronto, Canada*
smeshkani@ryerson.ca

***Bilal Farooq***
Laboratory of *Innovations in Transportation (LiTrans)*
*Ryerson University*
*Toronto, Canada*
bilal.farooq@ryerson.ca

***Abstract*— With the rapid increase in congestion, alternative solutions are needed to efficiently use the capacity of our existing networks. This paper focuses on exploring the emerging autonomous technologies for on-demand food delivery in congested urban cities. Three different last mile food delivery systems are proposed in this study employing aerial and ground autonomous vehicles technologies. The three proposed systems are: robot delivery system, drone delivery system and a hybrid delivery system. In the hybrid system the concept of hub-and-spoke network is explored in order to consolidate orders and reach more destinations in less time. To investigate the performance of the three proposed delivery systems, they are applied to the city of Mississauga network, in an in-house agent-based simulation in MATLAB. 18 Scenarios are tested differing in terms of demand and fleet size. The results show that the hybrid robot-drone delivery system performs the best with a fleet side of 25 robots and 15 drones and with an average preparation and delivery time less than the individual robot and drone system by 48% and 42% respectively.**

## I. Introduction

Travel demand in urban areas continues to grow resulting in increased congestion on the existing networks. As the population increases and the economy grows, car ownership continues to rise. For example, in Canada the number of vehicles registered increased by 1.8% from 2015 to 2016 and continued to increase by 1.6% from 2016 reaching 34.3 million vehicles in 2017 [1][2]. With this rate of increase, the existing infrastructure will not be able to keep up with the rise in demand and in fact, it is neither sustainable nor realistic to build enough roads and infrastructure to comfortably accommodate this increase. Therefore, there has been an ongoing movement across the globe to try and find solutions that are practical and feasible in terms of adapting to the growth in travel demand.

Parallel to that, on-demand services have shown large growth in recent years. A study by Castiglione et al. [3] has been conducted to evaluate and quantify the causes of congestion in San Francisco between 2010 and 2016 after congestion worsened. It was found that on-demand transportation companies such as Uber and Lyft are primary factors in contributing to the congestion. The effect of these services vary through the day, having the highest affect during evening peak hours, accounting for 70% of the increase in vehicle delay.

Adding to that in particular, food delivery Apps are reshaping how we purchase food. With the advancements in information and communication technologies, the traditional approach to ordering food over the phone and waiting for it to be delivered has been transformed into ordering digitally through a smartphone Apps and tracking one's order through the whole process up until the point of delivery or maybe even ordering and getting notified when the order is ready for pick-up if one chose that option. These services have their perks and have undeniably provided efficient on-demand last mile transportation services. Nevertheless, they are a major factor that contributes to the rise in congestion in urban areas. Therefore, this study explores solutions that can provide the same services, but without impacting the existing congestion on our road networks. Smart transportation models are emerging globally and automating last mile delivery services is being investigated to cope with the high increase in demand. By using the word automation two possibilities come in place; the use of autonomous ground vehicles (AGV's) and Autonomous aerial vehicles as a mean of transportation resulting in further redefining the traditional delivery services.

Autonomous ground vehicles (AGV's) is a generic term that represents autonomous cars that drive on the paved road or autonomous robots that operate on the sidewalk depending on the size, weight, and distance that the parcel needs to be delivered to. AGVs with parcel lockers will substitute current forms of regular parcel delivery, as current companies are testing this new delivery services of transporting goods for last mile delivery to provide a new level of convenience, and to reduce cost on the consumer but correspondingly increase companies' revenue [4]. The other type of AGV's that can be used are small autonomous vehicles that deliver parcels to the doorstep. These vehicles are relatively slower and use the sidewalk rather than the street to reach their destination. Hoffmann and Prause [5] discussed the regulatory framework for last mile-delivery robots by foreseeing the legal challenges these delivery robots will bring to the surface. However, no real case study was incorporated to systematically see the effect of applying these technologies on delivery timings or congestion and traffic.

On the other hand, and with the recent developments in technology, unnamed aerial vehicles or "drones" are being investigated for their potential to bring many changes in the way businesses and government agencies operate. Previously drones were being used mostly for military surveillance and tactical applications. However, currently, a wide range of applications for the use of drones in recreational and commercial purposes are being investigated. Drones are being utilized to transport packages, food or other goods. For example, drones can be used to provide health care services, postal delivery services as well as food delivery services.

Drones and robots are considered emerging technologies in the field of transportation and are defying the orthodox delivery systems currently used. Package companies and logistics firms are looking for the best options for that final step of a product's journey also known as last-mile, and an increasing number of companies are aiming to use robots and aerial drones to deliver goods to customers. McKinsey has predicted that autonomous vehicles including drones will deliver up to 80 percent of all items by 2025 [4].

Researchers have been exploring different aspects of drone delivery regarding its cost-effectiveness and impact on the environment but it's a fairly new research topic when it comes to autonomous deliveries. A study conducted by Jung and Kim [6] focused on analyzing specifications of amazon delivery drones and the delivery cost that accompanied with this new change, in using drones for the delivery. Though, delivery time and environmental impacts on drone delivery weren't considered in this research.

However, Goodchild and Toy [7] evaluated the effect of unmanned aerial vehicle technology in reducing CO2 emissions in the delivery service industry. The study concluded that net emissions differences can fluctuate vastly across delivery service zones, as they are profoundly influenced by the number of recipients and distance away from their take-off location. Yet, a life cycle assessment (LCA) should be pursued to acquire a more holistic perspective on environmental impacts.

The compelling need for service innovation has led to the development of this study. As demonstrated above each technology has been applied for use across various services. Nevertheless, these food delivery services share many similarities to customer transportation services. However, to the best of our knowledge there has not been any study evaluating the use of robots, drones, and their hybrid by a third-party delivery service. The premise is that these emerging technologies have the capability to relief congestion by taking on-demand delivery trips off the road network as well as delivery cost, while maintaining an equivalent or better service delivery times.

We propose a fully integrated delivery system for the on-demand last-mile delivery services that employ various automation technologies (ground and aerial) and is focused on taking the additional trips generated by such services off the already congested urban roads. Three delivery systems are investigated that are based on: a) robot technology based delivery system b) drone technology based delivery system and c) a hybrid of the two. A detailed analysis of system's operation and the integration of autonomous technologies in the delivery system is conducted in a real case study of city of Mississauga, Canada—a densely populated urban area. The case study focuses on the on-demand food delivery problem. The purpose of this study is multifold: a) investigation of the effectiveness of adding new autonomous technologies separately and combined to the on-demand food delivery system by maintaining a high level of service in terms of delivery times b) development of three different delivery models and c) comparison between them to determine the most efficient model in terms of minimum delivery times.

The remainder of this study is organized as follows. Section II explains the proposed methodology supported by a case study designed to explore different scenario of implementing autonomous technologies for delivery. Section III discusses the results after conducting a detailed analysis. Section IV presents conclusion and future work direction.

## II. Methodology

The performance of transportation systems is of crucial importance for the economic growth. When it comes to the last mile delivery in urban areas, efforts have been made to measure and control traffic movement within highly congested areas using conventional methods. However, with new technologies emerging this research will investigate the effectiveness of incorporating two emerging technologies; autonomous ground vehicles (robots) and autonomous aerial vehicles (drones) as alternative solutions for last mile food delivery. These systems are proposed as a means to providing the same quality of service that cars currently provide without using the road network thus reducing congestion.

### *A. General System Design*

Using the robot and drone technologies three types of delivery systems are designed and evaluated:

- The autonomous ground vehicle system. This system consists of a fleet of delivery robots (R) that operate on the sidewalk. This system assumes the robots travel at free flow speed with no pedestrian interference or waiting on signalized intersections for permission to cross. The robot system primary benefit is that it can navigate the sidewalks allowing it to bypass traffic easily
- The autonomous aerial vehicle system. This system will consist of a fleet of drones (D) that operate in a straight distance between two nodes assuming all high-rise buildings are at house level which allows drones to fly straight through. The drone system primary benefit is its ability to deliver at high speeds however there are strict government laws preventing the widespread use of commercial drones.
- A hybrid system utilizing both robots and drones. This system contains two phases. Phase one will have robots travel from the depot to the restaurants locations for pick-up and drop the package at the hub. Then phase two comes in, where the package is placed in a drone and sent from the depot to its drop-off destination. The hybrid system will utilize both technologies allowing robots to pick-up orders in highly densely populated areas where the landing of drones maybe difficult due to geographical conditions followed by drones to deliver food to farther locations where landing can be feasible.

### *A. Proposed delivery process for delivery robots and drone systems*

Two different transportation networks are developed and simulated in MATLAB. The robot system network N(I, L), consists of I intersections and L links representing roads. As for the drones system the network G(I, S) consists of I intersections and S Links representing the straight distance between two nodes assuming drones can fly without restrictions. Intersections I = {α, β} can be identified as either Restaurant (β) or Home (α).

Both systems consist of two agents: vehicles (vϵV) and food packages (pϵP). In the robot system vehicles are identified as robots (rϵR). In the drone system the vehicles are identified as drones (dϵD). Different fleet sizes are tested to determine the most efficient fleet size for three sets of demands (oϵO).

A centralized dispatching system (Di) to conduct the matching process is developed. When an order is placed (o), the dispatcher (Di) creates a food request table, which contains package ID, request time, Restaurant node, Home node. To account for meal preparation time the dispatcher adds 12 minutes to all orders requests time before assigning a vehicle to it. From every demand, a set of active orders requests are identified. Active orders (Ao) mean that the customer has placed an order and has not received it yet (Ao = {Ao_t, Ao_{t+1},…}).

After identifying the set of active order requests, every active request needs to be assigned to a vehicle. Therefore, the dispatcher (Di) checks the number of vehicles available every second during the simulation (t). Δt is the dispatch update time interval. The matching process policy of assigning a vehicle to an order is based on first-in first-out (FIFO). After determining the vehicle's availability, the dispatcher assigns the available vehicle to the first active order in queue based on a greedy algorithm assigning the vehicle available to the nearest restaurant.

Once the delivery has been dropped off to its destination, the vehicle becomes available and can be reassigned to the next order in queue in the food request table. If there are no orders to be assigned, the vehicle remains at the destination node waiting to be assigned when an order comes in queue. The steps used for both robot and drone system is described in detail in Algorithm 1 and 2. Different fleet sizes are explored to reach the optimal fleet size based on various sets of orders. The simulation keeps running until all the orders that have been placed in an hour have been delivered to the customers. The simulation involved following constraints:

- Every order had a minimum 12 min of preparation time
- Order capacity of a vehicle is 1
- For hybrid design orders are first consolidated at the depot from the restaurants and then delivered to the customers, no matter where the delivery location is
- Trips bunching is not allowed

### *B. Proposed delivery process for hybrid delivery system*

The major significant difference of this system is that it utilizes both robots and drones together operating from a hub/depot. There are two main benefits associated with such a hybrid system:

- one or more hubs can be strategically placed across the network where food from different restaurants can be consolidated together through a hub and be sent directly to their destinations. Such a system is called hub-and-spoke network and has been used by airline companies and in recently years by logistic companies to reduce transportation costs, improve cycle times and reduce inventory and more importantly serve more destinations for a given volume of output [8]. Having one or more hubs also allows but aerial and ground vehicles to recharge and be maintained while not in use, making this system a fully integrated system. Jung and Jaykrishnan [9] used the concept of hub-and-spoke network to increase the coverage point of electric shared taxis.
- Using both technologies at once allows the system to reap the benefits of these technologies since they possess different characteristics from their speeds to their navigation and trajectory systems. Both networks are used to operate this system and Intersections I = {α, β,μ} can be identified as either Restaurant (β), Home (α) or Depot (μ).

The system is divided into two phases; phase one uses robots to go from the depot (μ) to pick up the food package at the restaurant (β) and drops it off back at the depot (μ). The first phase was assigned to robots due to the robot characteristics of having lower speeds and operating on the sidewalks allowing the restaurant sufficient time to prepare the meal and therefore the dispatcher is not required to add 12 min to account for food preparation time. Phase two uses drones to carry the food package from the depot(μ) to the home (α) and heads back to the depot empty (μ). Phase two was chosen as drones because of their higher speeds and the fact that they need a designated landing area to place the food. For this system one centralized dispatcher is created. However, two food request tables are generated from it; robot food request table (ϒ) and drone food request table (τ).

Once a food order (o) is placed by a costumer, the order (o) is placed in the robot food request table(ϒ). An available robot (r_av) residing in the depot (μ) is assigned to pick-up the first package in queue from the restaurant (β) and bring it back to the depot (μ). After dropping off the food package (p) at the depot the order is assigned to the drone food request table (τ) with a time stamp equivalent to the robots drop-off time of the package. When an available drone is located in the depot(μ), it gets reassigned to the first package in queue and delivers the (p) from the depot (μ) to the home (α) thus completing the delivery of the order (o). The details for the hybrid system are described in Algorithm 3.

### *C. Performance criteria for all delivery systems*

To evaluate the performance of the three proposed systems, a Level of Service (LOS) scale is developed based on the customer average waiting time (w). Based on UberEATS business model identifying wait times for each restaurant from (15 minutes to 50 minutes) depending on the type of cuisine the restaurant services along with reference to the vehicle traffic Level of Service Categories and the criteria used in dividing those categories, a generic performance criterion in Table 1 was developed for all three systems. LOS should be evaluated for the overall delivery system and for each order separately to determine the optimal solution.

For the system overall LOS reached category implies the following: LOS A: Over reliable system; LOS B: Reasonably reliable system; LOS C: Stable system; LOS D: Approaching unreliable state system; and LOS F: Completely unreliable system.

Algorithm 1: Pseudo-code for Robots Food Delivery System

```
t=1;
Customer places order through the food delivery app;
Dispatcher (Di) identifies set of active requests Ao = {Ao_t, Ao_{t+1},…};
while active requests (Ao) exists do
    Dispatcher (Di) creates a food request table based on orders placed in a chronological
        order;
    Dispatcher (Di) starts searching for available robots (r_av) in the network;
    If available robots (r_av) are found
        Check all (r_av) distances from the first order in queue;
        Dispatcher (Di) assigns the first request lined up in queue to the nearest available
            vehicle
            to the restaurant for pick-up;
        The robot travels from its location to the assigned restaurant for pick-up
            using Dijkstra's shortest path algorithm;
        Dispatcher (Di) updates list of vehicles available and orders in queue;
    else
        Dispatcher (Di) waits until a robot in use becomes available to assign the food
package;
    end
    t=t+1;
end
```

Algorithm 2: Pseudo-code for Drones Food Delivery System

```
t=1;
Customer places order through the food delivery app;
Dispatcher (Di) identifies set of active requests Ao = {Ao_t, Ao_{t+1},…};
while active requests (Ao) exist do
    Dispatcher (Di) creates a food request table based on orders placed in a chronological
    order;
    Dispatcher (Di) starts searching for available drones in the network (d_av);
    If available drones (d_av) exist
        Check all (d_av) distances from the first order in queue;
        Dispatcher (Di) assigns the first request lined up in queue to the nearest available
            vehicle to the restaurant for pick-up;
        The drone travels from its location to the assigned restaurant for pick-up;
        Dispatcher (Di) updates list of drones available(d_av) and orders in queue;
    else
        Dispatcher (Di) waits until a drone in use becomes available (d_av) to assign the
            food Package to;
    end
    t=t+1;
end
```

Algorithm 3: Pseudo-code for Hybrid Food Delivery System

```
t=1;
Customer places order through the food delivery app;
Dispatcher (Di) identifies set of active requests Ao = {Ao_t, Ao_{t+1},…};
while active requests (Ao) exist do
    Dispatcher (Di) creates a robot food request table (ϒ) based on orders placed in a
        chronological order;
    Dispatcher (Di) creates a drone food request table (τ);
    if robot available (r_av) in depot (μ);
        Dispatcher (Di) assigns available robot (r_av) to the first active request lined up in
            queue (Ao_t) in (ϒ);
        (r) travels from depot (μ) to the assigned restaurant (β) for pick-up;
        (r) drops of (Ao_t) in depot (μ);
        Dispatcher (Di) updates list of robots available(r_av) at depot (μ);
        Dispatcher (Di) updates the set of orders (Ao_t) in (ϒ);
        when (Ao_t) reaches depot (μ) do;
            Dispatcher places (Ao_t) in queue in (τ);
            If drone available (d_av) in depot (μ);
                Dispatcher (Di) assigns available drone(d_av) to the first active request
                    lined up in queue (Ao_t) in (τ);
                (d) travels from its location to the assigned home (α) for drop-off;
                (d) drops-off (Ao_t) at home (α);
                (d) returns to depot (μ);
                Dispatcher (Di) updates list of drones available(d_av) at depot (μ);
                Dispatcher (Di) updates the set of orders (Ao_t) in (τ);
            else
                Dispatcher (Di) waits until a drone becomes available in the depot (μ);
            end
    else
        Dispatcher (Di) waits until a robot becomes available in depot (μ) to assign it to
            (Ao_t);
    end
    t=t+1;
end
```

TABLE 1 Level of Service Criteria

| LOS | Average Wait Time (min) |
|---|---|
| A | 1-20 |
| B | 21-30 |
| C | 31-40 |
| D | 41-50 |

## III. Case Study

For our case study the City of Mississauga in Ontario was chosen due to its highly dense populated urban area, being ranked the 6th most populated among cities and towns across Canada [10]. The road network was developed in MATLAB with an approximate area of (5.80 $km^2$) capturing the area as shown in Figure 1 with a total of 199 intersections and 286 streets links. The red dot represents the hub location for the hybrid system.

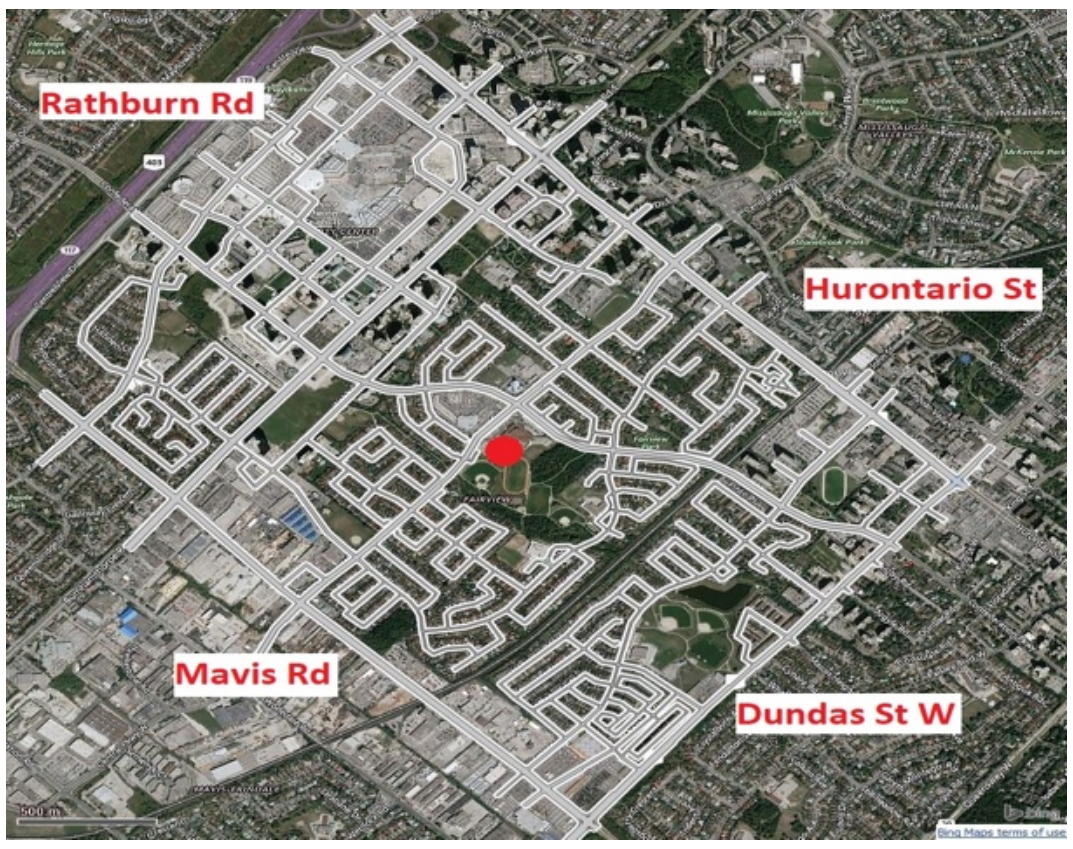

FIGURE 1 Mississauga network.

The area chosen is considered Mississauga city's downtown core consisting of both commercial and residential areas however the areas are not distributed uniformly. Downtown core consists of multiple small to medium commercial compounds as well as one major commercial compound known as Square One shopping Centre. Various restaurants exist within Square One shopping Centre and its surroundings. Furthermore, the area chosen has a high population density with multiple high-rise buildings due to its strategic location in downtown Mississauga.

In the simulation model nodes are identified as one of the following three: Restaurant, House, Depot. UberEats website is used to identify all the restaurants within the boundaries or surrounding the edges. The nearest node to the actual restaurant location is assigned as a restaurant node (origin nodes) and all the other nodes, which are not restaurants, were identified as house nodes (destination nodes). Furthermore, the location of depot is chosen based on the central location in the test network, where the number of nodes above the depot are approximately the same as below it, as well as on both sides from left and right. In this study due to small size of the studied network only one hub/depot is considered however the methodology is easily extendable to multiple hubs. In addition, the node assigned as depot is located near a park location so the design would be practical and there is sufficient space for an actual depot to be built. The study period for the simulation is during the evening peak period from 5:00 p.m. – 6:00 p.m. The demand used in this study is time dependent exogenous demand origin-destination matrices. From a set of restaurant nodes and house nodes, orders were created by matching a restaurant with a corresponding home node through uniform distribution creating an OD matrix. Time stamps assigned were uniformly distributed to each order between 5:00 p.m. – 6:00 p.m. During the study period, three demand scenarios were tested having the original demand set as 340 orders and increasing the demand by 10% & 20% growth rates. Moreover, different fleet sizes are tested for each delivery system. In total 18 scenarios are investigated as shown in Table 2 to compare the performance of the three proposed last mile delivery systems in terms of wait time and optimal fleet size required. We used larger fleet sizes for robot system than drones fleet sizes due to two primary factors:

- The robots have a speed of approximately 4.5m/s while the drones have a significantly higher speed of 40km/hr. The speed of robots was determined based on the speed of a Starships Company food delivery robot as for the drones the speed was set as half the maximum speed for an Amazon air prime delivery drone accounting for time to take off, landing time as well as the extra weight it may carry.
- Robots have to follow a designated road network while drones can fly straight distances

In the hybrid system we combine both robots and drone system and activate a depot location to develop a fully integrated autonomous delivery system that optimizes the benefits of using the two previous systems.

### *A. Food Delivery Systems Operating Policy Simulation*

The simulation time step is 1 second as well as the dispatcher update time interval. The simulation ends when all the food packages are dropped off at their destination and the food request table is cleared out.

For the robot and drone system, an example of a vehicle itinerary from the robot system with a fleet size of 75 robots operating at 20% increased demand set in Mississauga case study is presented in Table 3. Note that vehicle gets assigned after adding food preparation time. Where: o_ID: order ID; R_t: Order Request time (seconds); FP_t: Food Preparation Time (seconds); β: Origin node (Restaurant); α: Destination node (Home); P_t: Pick-up time from Restaurant (seconds) ; D_t: Drop-off time at Home (seconds) ; W_t: Customer wait time (seconds). An example of robot 1 path nodes that corresponds with order ID 374 is as follows: Vehicle(1).Pathnode=[190,189,154,155,160,163,166,167,120,121,124,125,126,127,133,132,131] For the robot routing system, it follows the Dijkstra's Shortest Path algorithm, and once the robot drops-off the order at the house node location it remains there until it gets reassigned to pick-up the next order. For the drones routing system, it follows a direct straight link connecting the nodes.

For the hybrid system, an example of an order going through the system with a robot fleet size of 25 and a drones fleet size of 15 operating at 20% increased demand set in Mississauga case study is presented in Table 4. Where: o_ID: order ID; R_t: Order Request time (seconds); β: Origin node (Restaurant); α: Destination node (Home); rP_t: Robot Pick-up time from Restaurant (seconds) ; rD_t: Robot Drop-off time at depot (seconds) ; dP_t: Drone Pick-up time from depot (seconds) ; dD_t: Drone Drop-off time at home (seconds) W_t: Customer wait time (seconds).

TABLE 2 List of Simulation Scenarios

| Scenario | System | Fleet Size | Demand |
|---|---|---|---|
| 1 | Robot | 25 robots | Original, 10%, 20% |
| 2 | Robot | 50 robots | Original, 10%, 20% |
| 3 | Robot | 75 robots | Original, 10%, 20% |
| 4 | Robot | 100 robots | Original, 10%, 20% |
| 5 | Robot | 125 robots | Original, 10%, 20% |
| 6 | Robot | 150 robots | Original, 10%, 20% |
| 7 | Drones | 5 drones | Original, 10%, 20% |
| 8 | Drones | 10 drones | Original, 10%, 20% |
| 9 | Drones | 15 drones | Original, 10%, 20% |
| 10 | Drones | 20 drones | Original, 10%, 20% |
| 11 | Drones | 25 drones | Original, 10%, 20% |
| 12 | Drones | 30 drones | Original, 10%, 20% |
| 13 | Hybrid | 25 robots–10 drones | 20% increase |
| 14 | Hybrid | 25 robots–15 drones | 20% increase |
| 15 | Hybrid | 25 robots–20 drones | 20% increase |
| 16 | Hybrid | 30 robots–10 drones | 20% increase |
| 17 | Hybrid | 30 robots–15 drones | 20% increase |
| 18 | Hybrid | 30 robots–20 drones | 20% increase |

TABLE 3 Sample of Robot (1) Itinerary

| o_ID | R_t | FP_t | β | α | P_t | D_t | W_t |
|---|---|---|---|---|---|---|---|
| 1 | 2 | 720 | 4 | 90 | 1204 | 1633 | 1631 |
| 91 | 830 | 720 | 126 | 56 | 2063 | 2319 | 1489 |
| 150 | 1263 | 720 | 148 | 113 | 2734 | 3311 | 2048 |
| 241 | 2111 | 720 | 10 | 190 | 3923 | 4776 | 2665 |
| 374 | 3293 | 720 | 167 | 131 | 5022 | 5377 | 2084 |

TABLE 4 Sample of Order (4) Itinerary

| o_ID | R_t | β | α | rP_t | rD_t | dP_t | dD_t | W_t |
|---|---|---|---|---|---|---|---|---|
| 114 | 1018 | 10 | 70 | 1532 | 1768 | 1768 | 1962 | 944 |

For the routing system used, since we are using both autonomous technologies; robots and drones, the robots phase uses the road network and follows the Dijkstra's Shortest Path algorithm, while the drones follow the straight trajectory between any two nodes therefore not needing to follow any routing algorithm

As can be seen from Table 4, robot pickup time was after 514 seconds from when the order was placed denoting that there were no available robots at the time the request was made. The order gets placed in queue for an available robot to return to the depot. Moving on to phase two, since there was an available drone at the depot the order got picked-up immediately after being dropped-off at the depot and was dropped-off to the client totaling up the costumers wait time to 944 seconds.

## IV. Results & analysis

Figure 2 and 3 present the average wait time for robot and drone systems according to different demand scenarios. The results show that for the robot and drone system, when increasing the fleet sizes it reduces the average wait time for customers which is an apparent conclusion because when we have more vehicles in the system the better the performance of the system will be however, after increasing the fleet size the average waiting time plateaus. For the robot system the wait time starts to level around 100 robots as for the drone system it starts to plateau after 15 drones reaching a level state from 20 drones onwards for our three sets of demand. We can determine from Figure 2 and 3 that the optimal fleet size for our demand scenarios for robots and drones are; 75 robots and 10 drones with an average wait time for the maximum demand being 32.4 minutes and 28.8 minutes respectively.

However, there is an approximate 87% difference in optimal fleet sizes between the robot system and the drone system, which results in an 11% difference in the average wait time due to the difference in speed between robots and drones. Therefore, the hybrid solution was introduced to maximize the benefits of both autonomous technologies an activate a depot placed at a strategic location to accommodate for the uneven distribution of commercial and residential areas within our study area.

As shown in Figure 4, for the hybrid delivery system, six different scenarios are analyzed. From Figure 4 it can be seen that wait time fluctuated between the six scenarios however, one thing stood out was the number of drones played a significant role in increasing or decreasing the average wait time due to the drone's capability of delivering the food at significantly higher speeds. Therefore, adding 5 drones with keeping 25 robots reduced the average delivery time by approximately 50%. The optimal scenario was having 25 robots and 15 drones. While this hybrid system has an overall LOS C, stable system, based on the average wait time, it is not sufficient to look at only the LOS of the system. The LOS of each order should be assessed because from the average wait time it appeared that 25 robots and 10 drones provides a stable system but in fact there was a high percent of orders that fell in the LOS F category as displayed in Figure 5, meaning some orders had to wait a long time in queue until a drone became available.

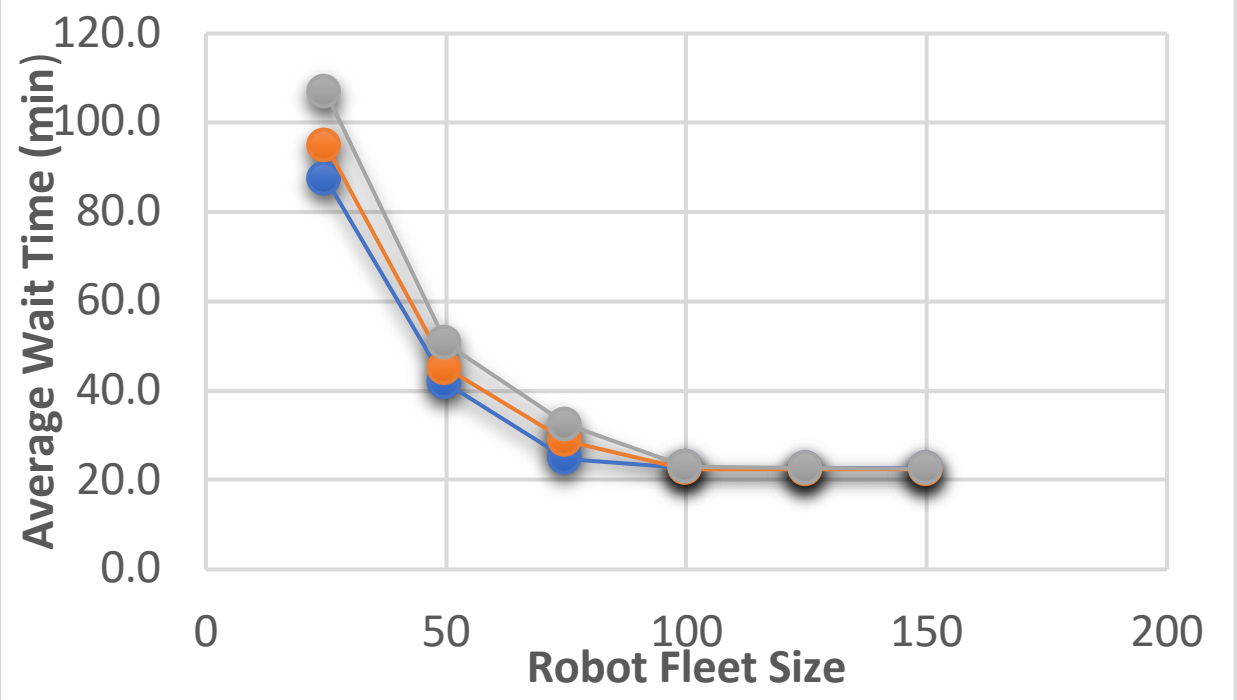

FIGURE 2 Average customer wait time for robots system.

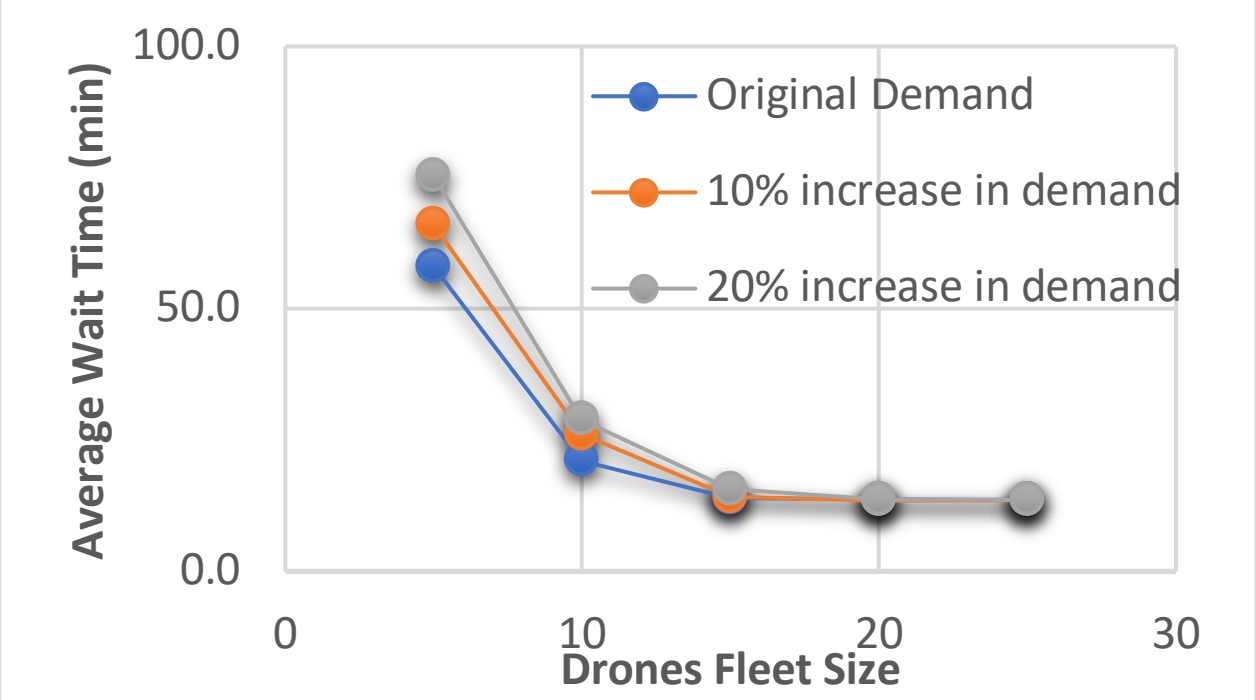

FIGURE 3 Average customer wait time for drones system.

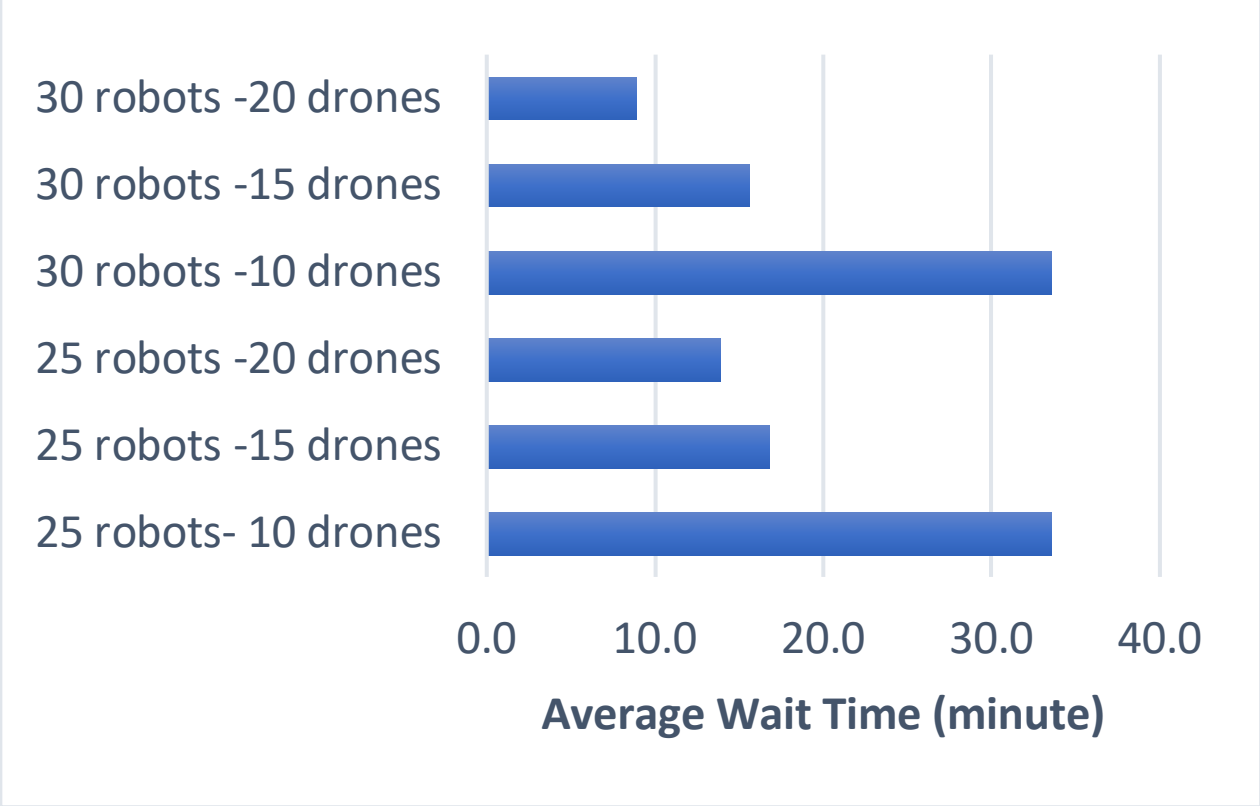

FIGURE 4 Average wait time for hybrid system.

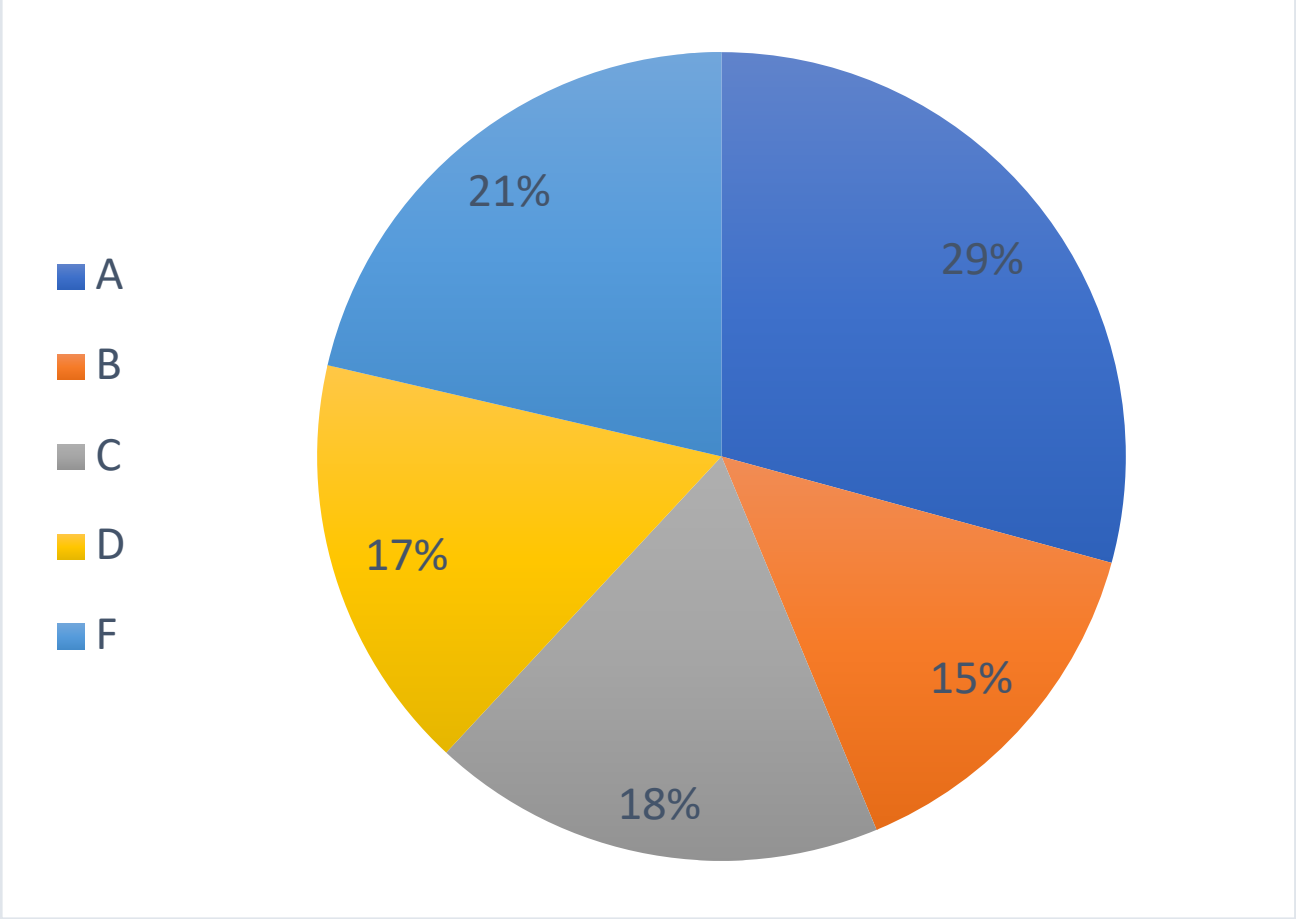

FIGURE 5 Level of service for scenario 25 Robots - 10 drones in the hybrid system.

To compare all three system, w box plot is developed (refer to Figure 6) t with the optimal fleet sizes determined previously per system. As demonstrated the maximum wait time for the optimal robot fleet size is 50 minutes, according to UberEATS website delivery time is usually set between 45-50 minutes per order. However, when applying the drone system, the maximum delivery times reduces by approximately 12%. Furthermore, when applying the hybrid system, it reduces the maximum wait time from the robot system by 43% and from the drone system by 34% making the hybrid system the most efficient of all three systems. However, when analyzing the minimum wait time for the hybrid system its 2.5 minutes, which may not be sufficient to some restaurants that are located closer to the depot to prepare the food therefore food preparation time should be allocated to the restaurants that are located near the hub before assigning a robot for pick-up. When analyzing the median waiting time, it is observed that for robots and drones median waiting times are less than 10% apart compared to each other. However, when comparing the median waiting time for the robot system with the hybrid system, the waiting time is reduced by 46% under hybrid scenario. For the robot and drone systems 50% of the orders are delivered above (30-33) minutes whereas for the hybrid system 50% of the orders are delivered more than 17 minutes and up to 29 minutes.

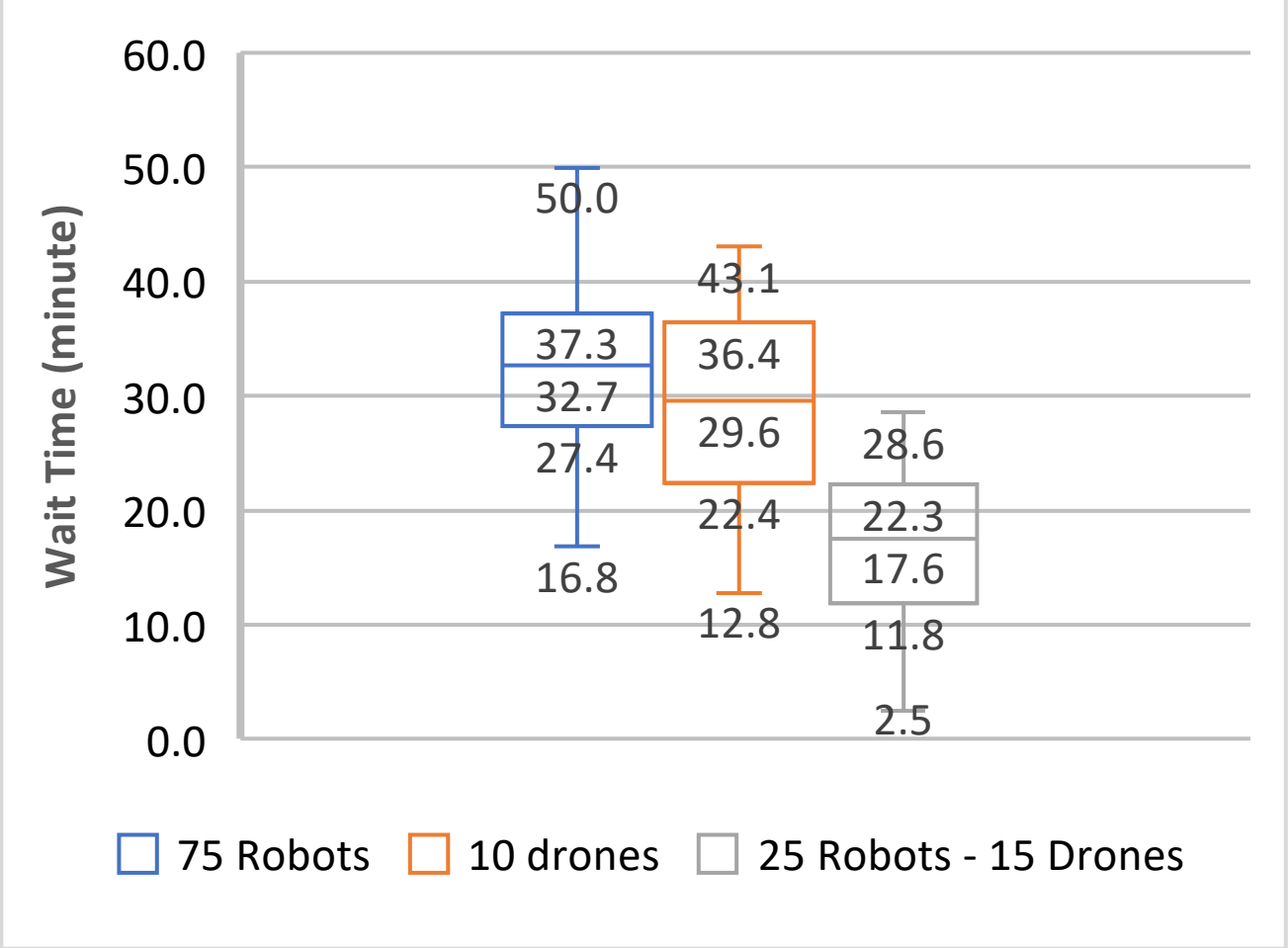

FIGURE 6 Average wait time for optimal fleet sizes per system.

## V. Conclusions

In this study three innovative last mile food delivery systems are proposed employing aerial and ground autonomous vehicles in order to mitigate traffic congestion caused by new and upcoming on-demand food delivery apps. Three systems proposed are: robot-based delivery system, drone-based delivery system and a hybrid system integrating robots and drones.

To test the efficiency of the three proposed systems, they were applied to city of Mississauga in an in-house developed simulation platform during evening peak period. In total 18 scenarios were tested where demand and fleet size were systematically increased. Average wait time and optimal fleet size were used as performance measure. The results obtained from simulation showed that for the robot system, a large fleet size is needed to accommodate the applied demand, since the robots operate on the sidewalks and on a lower speed. As for the drones, while a significantly lower fleet size (86% less than robots) was used, there are strict regulations on operating drones for commercial use. Furthermore, with the existence of high-rise buildings processionary measures need to be made for landing as well as adjustments to the drone's trajectory.

For the hybrid system, both technologies were incorporated in the system and a strategic depot location was identified making a hub-and-spoke network. Robots were chosen for phase one of the delivery due to their slower speed, giving time to the restaurant for food preparation, followed by phase two which used drones to drop-off the food The hybrid system was found to be the most efficient of the three systems since it accommodates for the food preparation phase as well as has a depot location that can be used for storage and charging purposes as well as distribution center.

The results showed that at maximum operating demand the hybrid system with optimal fleet size had the least average delivery time being 48% less than the optimal fleet size of robots and 42 % less than the optimal fleet size of drones, with the least variance between orders delivery times.

For future research, a number of actions can be taken to further develop these systems. In current design, customers are served based on (FIFO). However, for the future policies an algorithm can be developed to optimize the assignment of vehicles to orders. Furthermore, to develop a fully integrated system, autonomous vehicles can be added to these systems and can account for delivery sizes that cannot be carried by robots or drones (e.g. catering services). Moreover, the current three systems proposed carry one order at a time. However, for future purposes these systems can be upgraded so vehicles can pick-up multiple packages and drop them off at once. Finally, a p-hub optimization algorithm can be used to determine the optimal depot location to minimize customer wait time, optimize charge time of the robots and drones, while maximizing companies profit. In real applications, the robot will interact with pedestrians and e-scooters on the sidewalk. Furthermore, in the intersections interactions with vehicles and bicycles are expected. A detailed health and safety analysis would be required for such situations.